\documentclass[12pt]{article}
\usepackage{amssymb}
\usepackage{amsmath}
\usepackage{amsthm}
\usepackage{epsfig}
\usepackage[mathscr]{eucal}
\def\RR{{\mathbb R}}

\def\proof{{\bf Proof: \, }}
\def\beqns{\begin{eqnarray*}}
\def\eeqns{\end{eqnarray*}}
\def\beqn{\begin{eqnarray}}
\def\eeqn{\end{eqnarray}}

\def\endproof{{\hfill $\blacksquare$}}
\def\mendproof{{\qquad \blacksquare}}

\def\dis{\displaystyle}
\def\sph{S^{d-1}}
\def\no{\noindent}
\def\no{\noindent}
\def\la{\langle}
\def\ra{\rangle}

\def\dis{\displaystyle}
\def\sph{S^{d-1}}

\def\no{\noindent}
\def\no{\noindent}
\def\la{\langle}
\def\ra{\rangle}

\usepackage{amssymb,amsmath,amsthm}

\numberwithin{equation}{section}

\newtheorem{theorem}{Theorem}
\newtheorem{lemma}[theorem]{Lemma}

\theoremstyle{remark}

\begin{document}

\title{On Separation of  Minimal Riesz Energy
Points on  Spheres in Euclidean Spaces}

\author{A. B. J. Kuijlaars\thanks{The first author is supported in part by FWO-Flanders
projects G.0176.02 and G.0455.04, by K.U.Leuven research grant
OT/04/24, by INTAS Research Network NeCCA 03-51-6637, and by NATO
Collaborative Linkage Grant PST.CLG.979738.}, E. B.
Saff\footnote{This author is supported, in part, by a U.S.
National Science Foundation grant DMS-0296026.}, and X. Sun}

\maketitle

\abstract{Let $S^d$ denote the unit sphere in the Euclidean space $\RR^{d+1}\; (d \ge 1)$.
Let $N$ be a natural number $(N \ge 2)$, and let $\omega_N:=\{x_{1},
 \ldots,x_{N}\}$ be a collection of $N$ distinct points on $S^d$ on which the minimal
 Riesz $s$-energy is attained. In this paper, we show that the  points $x_{1},
 \ldots,x_{N}$ are well-separated for the cases $d-1 \le s <d$.}

\medskip

\def\dis{\displaystyle}
\def\sph{S^{d-1}}
\def\no{\noindent}
\def\no{\noindent}
\def\la{\langle}
\def\ra{\rangle}

\medskip

\section{ Introduction.} \label{intro}
\medskip

Let $S^d$ denote the unit sphere in the Euclidean space $\RR^{d+1}\; (d \ge 1)$.
Let $N$ be a natural number $(N \ge 2)$, and let $\omega_N:=\{x_{1},
 \ldots,x_{N}\}$ be a collection of $N$ distinct points on $S^d$. The Riesz
$s$-energy ($s \ge 0$) associated with $\omega_N$,
$E_s(\omega_N)$, is defined by
$$E_s(\omega_N):=
\begin{cases}
{\dis \sum_{i \ne j}{1 \over |x_i-x_j|^{s} }}, & \text{if}  \quad s>0, \\
 {\dis  \sum_{i \ne j}\log{1 \over  |x_i-x_j| }},          & \text{if} \quad s=0. \end{cases}$$
Here $|\cdot|$ denotes the Euclidean norm. We use ${\cal E}_s(S^d,
N)$ to denote the {\bf $N$-point minimal $s$-energy } over $S^d$
defined by \beqn {\cal E}_s(S^d,N):=\min_{\omega_N \subset
S^d}E_s(\omega_N), \label{mini} \eeqn where the infimum is taken
over all $N$-point subsets of $S^d$. If $\omega_N \subset S^d$ is
such that
$$ E_s(\omega_N)={\cal E}_s(S^d,N),$$
then $\omega_N $ is called a minimal $s$-energy configuration, and
the points in $\omega_N $ are called minimal $s$-energy points, or
simply minimal energy points if the linkage to the parameter $s$
is well-understood in a certain context. It is obvious that
minimal $s$-energy configurations exist. Also, if $\omega_N $ is a
minimal $s$-energy configuration, and if $\rho$ is a metric space
isometry from $S^d$ to $S^d$, then the image of $\omega_N $ under
$\rho$, $\rho(\omega_N)$, is also a minimal $s$-energy
configuration. Minimal $(d-1)$-energy points are often referred to
as ``Fekete points"; see \cite{dahlberg}. The determination of
minimal $s$-energy configurations and the corresponding minimal
$s$-energy on $S^d$ and other manifolds is an important problem
that has applications in many subjects including physics,
chemistry, computer science, and mathematics. For further background
regarding this problem and its
applications, we refer readers to the  expository papers by Hardin
and Saff \cite{hs2}, and by Saff and Kuijlaars \cite{ks2}. The
papers \cite{dg}, \cite{gotz1}, \cite{hs1}, \cite{ks1}, and \cite{rsz2} and the
references therein also contain valuable pertinent information.

The determination of the distribution of minimal $s$-energy points in the cases $d\ge 2$ turns
out to be rather elusive. It is, however, generally expected  that these points are ``well-separated"
in the sense that there exists a positive constant $A_{d,s}$,
 depending  only on $d$ and $s$, such that
 \beqn \min_{i \ne j}|x_i -x_j| \ge A_{d,s}N^{-1/d}. \label{separation} \eeqn
Dahlberg~\cite{dahlberg} proved that Fekete points are well-separated.
(Dahlberg~\cite{dahlberg} actually established the well-separatedness
of Fekete points on every compact $C^{1, \alpha}$-surface in $\RR^{d+1}$).
Kuijlaars and Saff \cite{ks1} proved that minimal $s$-energy points are
well-separated for the cases $s >d$. There have been a series of more quantitative results
regarding the
case $d=2, s=0$, which corresponds to the minimal logarithmic energy on $S^2$.
Rakhmanov, Saff, and Zhou \cite{rsz} first showed that
$3/5$ is a lower bound for the constant $A_{2,0}$ in inequality~(\ref{separation}).
Dubickas \cite{dubi} refined Rakhmanov, Saff, and Zhou's method, and showed that
$7/4$ is a lower bound for the constant $A_{2,0}$ in inequality~(\ref{separation}). Using potential theory
and stereographical projection techniques, Dragnev \cite{drag} established an appealing
lower bound for the minimum separation of the minimal logarithmic energy points on $S^2$ to be $2(N-1)^{-1/2}$.

 In this paper, we show that the minimal $s$-energy points are
well-separated for the cases $d-1 \le s <d$. Note that the case $s=d-1$ is already
covered by the aforementioned Dahlberg's result. While the parameter $s$ is restricted in the
range $d-1 \le s <d$, the outcome $s=0$ can only occur when $d=1$, putting the problem
on the unit circle $S^1$. It is shown by G\"{o}tz \cite{gotz2} that  the $N$th roots of unity
and their
rotations are the only minimal $s$-energy configurations on $S^1$;
see also \cite{mmrs}. Therefore in this paper, we can use the tacit assumption that $s>0$. Our proof entails
comparing the Riesz $s$-potentials on the slightly larger sphere of radius $1+N^{-1/d}$
of two probability measures: the rotationally invariant probability measure on $S^d$,
and the normalized counting measure on a minimal $s$-energy
configuration $\omega_N$. The Riesz $s$-potential of the rotationally invariant probability measure on $S^d$
can be  expressed in closed form in terms of the Gauss hypergeometric functions $\,_2F_1(a,b;c;z)$.
The crux of our argument is an application of the principle of domination for
$\alpha$-superharmonic functions; see \cite{landkof}. This paper is organized
as follows. In Section 2, we introduce the necessary notations and terminologies.
Also in Section 2, we list a few formulas pertaining to the
hypergeometric functions that we will use in our proofs. In Section 3, we state and prove our
main result.

\section{ Notation and terminology.} \label{notation}
\medskip

Given a minimal $s$-energy configuration $\omega_N$, we use $v_N$ to
denote the normalized counting measure on $\omega_N$, i.e.,
$$v_N:=N^{-1}\sum^N_{j=1}\delta_{x_j},$$
where $\delta_{x_j}$ denotes the unit point mass at $x_j$.
The rotationally invariant probability measure on $S^d$ is denoted by $\mu$.
For a $\sigma$-finite positive Borel measure $\lambda$ supported on a compact subset
$K$ of $\RR^{d+1}$, we define its Riesz $s$-potential
$U_s^\lambda \;(s>0)$  by
$$U_s^\lambda (x):= \int_K |x-y|^{-s}d\lambda(y).$$
Note that $U_s^\lambda$ may take the extended value $\infty$ on some subsets of $\RR^{d+1}$.
In this paper, $K$ is either $S^d$ or $\omega_N$.
In the next section, we will show that the  Riesz $s$-potential of
$\mu$ can be expressed in closed form in terms of the Gauss hypergeometric functions $\,_2F_1(a,b;c;z)$
defined by
$$\,_2F_1(a,b;c;z):=\sum^\infty_{n=0}{(a)_n (b)_n \over (c)_n}{z^n \over n!},$$
where $(a)_n$ is the Pochhammer symbol defined by
$$(a)_n:=
\begin{cases}
1, & \text{if} \quad  n=0, \\
               a(a+1)\cdots (a+n-1), & \text{if} \quad n \ge 1. \end{cases}$$
There are several sources that provide essential properties of
the hypergeometric functions $\,_2F_1(a,b;c;z)$ such as Abramowitz et al \cite{ambra},
and Andrews et al \cite{askey}.
We will be using a few basic formulas
pertaining to the hypergeometric functions $\,_2F_1(a,b;c;z)$, which can all be  found in \cite{ambra}.
We quote them here for easy reference.
Under the conditions ${\rm Re}\,(c)>{\rm  Re}\,( b)>0$,  the following formula holds true:
\beqn \int_0^1(1-zu)^{-a}
       u^{b-1}   (1-u)^{c-b-1}du
       ={\Gamma(b)\Gamma(c-b)\over \Gamma(c)}
       \,_2F_1(a,b;c;z).   \label{euler} \eeqn
The above integral represents an analytic function in the $z$-plane
cut along the real axis from $1$ to $\infty$. Formula~(\ref{euler}) is often
called Euler's integral representation for the hypergeometric function $_2F_1$.
If ${\rm Re}(c-a-b)<0$, then
\beqn \lim_{z \rightarrow 1^{-}}{\,_2F_1(a,b;c;z)\over (1-z)^{c-a-b}}
={\Gamma(c)\Gamma(a+b-c)\over \Gamma(a)\Gamma(b)}. \label{hyperlim1} \eeqn
If ${\rm Re}(c-a-b)>0$, then
\beqn \,_2F_1(a,b;c;1)
={\Gamma(c)\Gamma(c-a-b)\over \Gamma(c-a)\Gamma(c-b)}. \label{hyperlim2} \eeqn
The following derivative formula can be easily proved by term-by-term differentiation in a suitable domain:

\beqn {d \over dz}\,_2F_1(a,b;c;z)={ab \over c} \,_2F_1(a+1,b+1;c+1;z). \label{hyperderiv} \eeqn

\medskip
\section{ Main Result and  Proofs.}\label{result}
\medskip
On various occasions, we use $C_{d,s}$ to denote some unspecified positive constants,
depending only on $d$ and $s$. The exact values of $C_{d,s}$ may be different from
proof to proof. In the same proof, however, for clarity we use  different
notations for different constants, namely $C'_{d,s}$, and $C''_{d,s}$ if necessary.
Although in the current paper, we  do not strive to
estimate these constants, closed forms for them can be obtained with
some devoted calculations. We will be using the notations  $\sum_{i\ne j}$, and
$\sum_{j:j\ne i}$. The former denotes a ``double sum", excluding only those terms
with $i=j$. The latter denotes a single sum in which $i$ is fixed, and the summation is done
only on $j$.

The following lemma is well-known; see e.g. \cite{ks1}.
\begin{lemma} \label{le1}
 For $d-1 \le s<d$, there exists a positive constant $C_{d,s}$ independent of $N$, such that
 for any $N$ distinct points $x_1, \ldots, x_N$ on $S^d$, we have
\beqn \sum_{i\ne j}|x_i-x_j|^{-s} > \gamma_{d,s} N^2 -C_{d,s} N^{1+s/d}, \label{energy} \eeqn
where
$$\gamma_{d,s}: =\int_{S^d} \int_{S^d}|x-y|^{-s}d \mu(x)d \mu(y)
={\Gamma((d+1)/2)\Gamma(d-s) \over \Gamma((d-s+1)/2)\Gamma(d-s/2)}.$$
\end{lemma}
\medskip

By inspecting the pertinent proof in \cite{ks1}, we find that a
quantitative estimate of the constant $C_{d,s}$ in the above lemma
is possible. The determination of asymptotically sharp values for these
constants has received much attention in the literature; see
\cite{b}, \cite{ks1}, \cite{rsz}, and references therein.

\begin{lemma} \label{kuijlaars}
 For $0 < s<d$, there exists a positive constant $C_{d,s}$ independent of $N$, such that
\beqn U_s^{v_N}(x) \ge \gamma_{d,s} - C_{d,s}N^{-1+s/d}, \quad |x|=1. \label{inequality1} \eeqn
\end{lemma}

\proof
Since $v_N$ is the normalized counting measure of a  minimal $s$-energy
configuration $\omega_N = \{x_{1},
 \ldots,x_{N}\}$, we have for each fixed $i, \quad 1 \le i \le N$,
$$\sum_{j:j\neq i}|x-x_j|^{-s} \ge \sum_{j: j\neq i}|x_i-x_j|^{-s}, \quad x \in S^d.$$
Summing over $i$ and using  Lemma~\ref{le1}, we get
\beqn
(N-1)\sum^N_{j=1}|x-x_j|^{-s} \ge  \sum_{i\neq j}|x_i-x_j|^{-s}
\ge \gamma_{d,s} N^2 -C_{d,s} N^{1+s/d}. \label{ineq2}
\eeqn
Dividing by $N(N-1)$ we get the desired estimate. \endproof

\medskip
\begin{lemma} \label{le2}
For each fixed $s>0$,
the potential $U_s^\mu$ of the measure $\mu$ is a radial function, and
has the explicit expression in terms of the hypergeometric
function:
\beqns U_s^\mu(x)
       &=  & ( R+1)^{-s}\,_2F_1\left({s\over 2},{d\over 2};d;{4R \over (R+1)^2}\right), \quad |x|=R\ne 1. \eeqns
       \end{lemma}

\proof
For $x\in \RR^{d+1},\;|x|\ne 1, $ we have
$$U_s^\mu(x)=\int_{S^d}|x-y|^{-s}d\mu(y).$$
Let $x,y \in \RR^{d+1}$ with $|x|=R$ and $|y|=1$. Denote the angle between the
two vectors $x$ and $y$ by $\theta$. Then $\cos \theta =\la {x\over R}, y\ra$.
By the law of cosine, $|x-y|^2=R^2+1-2R\la {x\over R}, y\ra$.
Thus by using the Funk-Hecke formula; see M\"{u}ller \cite{muller}, we have
\beqns \int_{S^d}|x-y|^{-s} d\mu(y)
       &   =&\int_{S^d}(R^2+1-2R\la {x\over R}, y\ra)^{-s/2} d\mu(y)\\
        & =&{\nu_{d-1}\over \nu_{d}} \int_{-1}^1(R^2+1-2Rt)^{-s/2} (1-t^2)^{(d-2)/2}dt, \eeqns
where $\nu_d$ denotes the surface area of $S^d$.
Using the substitution $2u=t+1$ and
Euler's integral representation of the hypergeometric function $_2F_1$, we have
\beqns     U_s^\mu(x)
               &=     &2^{d-1}(R+1)^{-s}{\nu_{d-1}\over \nu_{d}}
                \int_0^1\left(1-{4R\over (R+1)^2}u\right)^{-s/2}
       u^{(d-2)/2}   (1-u)^{(d-2)/2}du\\
      & =&2^{d-1}(R+1)^{-s}{\nu_{d-1}\over \nu_{d}}{\Gamma^2({d\over 2})\over \Gamma(d)}
       \,_2F_1\left({s\over 2},{d\over 2};d;{4R \over (R+1)^2}\right). \eeqns
To simplify, we use the formula
 (see \cite{muller}) $$\nu_d={2 \pi^{(d+1)/2} \over \Gamma({d+1 \over 2})},$$
and then the formula (see \cite{ambra})
 \beqn \Gamma(2z)=(2\pi)^{-1/2}2^{2z-1/2} \Gamma(z)\Gamma(z+1/2),\label{double}\eeqn
with $z = d/2$. We have
\beqns   U_s^\mu(x)
                 &=&2^{d-1}(R+1)^{-s}{\Gamma({d+1 \over 2}) \over \sqrt{\pi}
                \Gamma({d \over 2})}{\Gamma^2({d\over 2})\over \Gamma(d)}
       \,_2F_1\left({s\over 2},{d\over 2};d;{4R \over (R+1)^2}\right) \\
       &=&(R+1)^{-s}
       \,_2F_1\left({s\over 2},{d\over 2};d;{4R \over (R+1)^2}\right).  \qquad  \mendproof \eeqns

The following two special cases of Lemma~\ref{le2} are worth noting. Firstly, when $0<s<d$,
the potential $U_s^\mu(x)$ is well defined for $|x|=1$. In fact, a simple application
of the Lebesgue Dominated Convergence Theorem yields
\beqns  \lim_{|x| \rightarrow 1}U_s^\mu(x)
       &=  &\lim_{|x| \rightarrow 1}\int_{S^d}|x-y|^{-s}d\mu(y)
       =\gamma_{d,s}. \eeqns
On the other hand, when $0<s<d$, the hypergeometric
function
$\,_2F_1\left({s\over 2},{d\over 2};d;z\right)$ is continuous at $z=1$. Using equation~(\ref{hyperlim2}), we
have
\beqns \lim_{R \rightarrow 1}(R+1)^{-s}
       \,_2F_1\left({s\over 2},{d\over 2};d;{4R \over (R+1)^2}\right)
       &=&2^{-s}\,_2F_1\left({s\over 2},{d\over 2};d;1\right) \\
       &=&2^{-s}{\Gamma(d) \Gamma({d-s \over 2})\over \Gamma(d/2) \Gamma(d-s/ 2)}. \eeqns
Thus, we have
\beqn \gamma_{d,s}=2^{-s}{\Gamma(d) \Gamma({d-s \over 2})\over \Gamma(d/2) \Gamma(d-s/ 2)}.
\label{twocon} \eeqn

Due to the importance of the constant $\gamma_{d,s}$ in this paper, we
feel reassured that we are  able to verify equation~(\ref{twocon}) directly, and we
share the reassurance with the readers.
Using
equation~(\ref{double}) with $z=d/2$, we write
$$\Gamma(d)=(2\pi)^{-1/2}2^{d-1/2} \Gamma(d/2)\Gamma((d+1)/2),$$ we have
\beqns
2^{-s}{\Gamma(d) \Gamma({d-s \over 2})\over \Gamma(d/2) \Gamma(d-s/ 2)}
&= &2^{-s}{(2\pi)^{-1/2}2^{d-1/2} \Gamma(d/2)\Gamma((d+1)/2)\Gamma({d-s \over 2})\over \Gamma(d/2) \Gamma(d-s/ 2)}\\
&= &{(2\pi)^{-1/2}2^{d-s-1/2} \Gamma({d-s \over 2})\Gamma({d-s+1 \over 2})\Gamma((d+1)/2)\over
\Gamma({d-s+1\over 2}) \Gamma(d-s/ 2)}\\
&= &{ \Gamma(d-s )\Gamma((d+1)/2)\over
\Gamma({d-s+1\over 2}) \Gamma(d-s/ 2)}.
\eeqns
Here in the last step, we have used equation~(\ref{double}) again with $z=(d-s)/2$.
Secondly, when $d=2$, the potential $U_s^\mu(x)$ has the elementary form:
$$U_s^\mu(x)={1 \over 2R}{(1+R)^{2-s}-|R-1|^{2-s} \over 2-s}, \quad |x|=R, \quad s \ne 2.$$

\medskip
\begin{lemma} \label{le3}
Assume $d-1\le s <d$. Then there exists a positive constant $C_{d,s}$, independent of $N$, such that
$$U_s^\mu(x)>\gamma_{d,s} -C_{d,s}N^{-1+s/d}, $$
for all $x \in \RR^{d+1}$ with $|x|=1+N^{-1/d}$.
\end{lemma}
\proof We first note that with $R_{N,d}:=1+N^{-1/d}$, we have \beqn
(R_{N,d}+1)^{-s} = 2^{-s} (1- \frac{s}{2}
N^{-1/d})+o(N^{-1/d}). \label{ds1} \eeqn We now estimate
      $\,_2F_1\left({s\over 2},{d\over 2};d;{4R_{N,d} \over (R_{N,d}+1)^2}\right)$.
      By using the Fundamental Theorem of Calculus and then equation~(\ref{hyperderiv}), we have
\beqns {\lefteqn{ \,_2F_1\left({s\over 2},{d\over 2};d;{4R_{N,d} \over (R_{N,d}+1)^2}\right)}} \\
      & = &\,_2F_1\left({s\over 2},{d\over 2};d;1\right) -
      \left[\,_2F_1\left({s\over 2},{d\over 2};d;1\right)-
           \,_2F_1\left({s\over 2},{d\over 2};d;{4R_{N,d} \over (R_{N,d}+1)^2}\right)\right] \\
      & = & {\Gamma(d) \Gamma({d-s \over 2})\over \Gamma(d/2) \Gamma(d-s/ 2)}  -
            {s \over 4} \int^1_{4R_{N,d}/(R_{N,d}+1)^2}\,_2F_1\left({s\over 2}+1,{d\over 2}+1;d+1;z\right)dz.
            \eeqns
We use  equation~(\ref{hyperlim1}) to estimate the above integral.
For any given $\epsilon >0$, we have, for $N$ sufficiently large,
that
$$\left|{\,_2F_1\left({s\over 2}+1,{d\over 2}+1;d+1;z\right) \over
(1-z)^{(d-s-2)/2}}-\beta_{d,s}\right| < \epsilon,\quad z \in [4R_{N,d}/(R_{N,d}+1)^2, 1],$$ where
$$\beta_{d,s}:={\Gamma(d+1) \Gamma((d-s)/2+1)\over \Gamma(d/2+1) \Gamma(s/2+1)}.$$
This implies
\beqn {\lefteqn{ \int^1_{4R_{N,d}/(R_{N,d}+1)^2}\,_2F_1\left({s\over 2}+1,{d\over 2}+1;d+1;z\right)dz}}  \nonumber\\
      &\le &(\beta_{d,s}+\epsilon)
      \int^1_{4R_{N,d}/(R_{N,d}+1)^2}(1-z)^{(d-s-2)/2} dz \nonumber\\
     &= &(\beta_{d,s}+\epsilon)
      {2 \over d-s} \left({R_{N,d}-1\over R_{N,d}+1}\right)^{d-s}\nonumber\\
    &\le & {2^{s-d+1} \over d-s}(\beta_{d,s}+\epsilon)
       N^{-1+s/d}. \label{ds2}
           \eeqn
Combining Lemma 3 and the above two estimates (\ref{ds1}) and (\ref{ds2}), and noting that
$1/d \ge 1-s/d$, we have for $x$ with $|x|=R_{N,d}$,
\beqns  U_s^\mu(x)
       & =& 2^{-s} (1-\frac{s}{2} N^{-1/d})\left({\Gamma(d) \Gamma({d-s \over 2})\over \Gamma(d/2) \Gamma(d-s/ 2)}
       -C_{d,s} N^{-1+s/d}\right)+o(N^{-1+s/d}), \eeqns
which gives the desired result of Lemma~\ref{le3}.
\endproof

In the discussion that follows, we will need the notion of ``$\alpha$-superharmonic functions" and the principal of
domination of $\alpha$-superharmonic functions. These topics can be found in Landkof~\cite{landkof}.
The definition of $\alpha$-superharmonic functions is technical. Upon checking the pertinent
material in Landkof~\cite{landkof}, one finds  the relation $d+1-\alpha=s$ between the parameter $\alpha$
used in Landkof~\cite{landkof} and the parameter $s$ we use here. Thus the requirement $d-1 \le s <d$ translates
into $1 < \alpha \le 2$ in Landkof~\cite{landkof}. Furthermore, what is meant in Landkof~\cite{landkof}
by an $\alpha$-superharmonic function is in fact a $(d+1-s)$-superharmonic function here in our context.
We choose not to use the phrase $(d+1-s)$-superharmonic function because we feel that notion
$\alpha$-superharmonic function has been coined in the mathematical literature.
In Chapter 1,
Section 5 of  \cite{landkof}, it is proved that for a $\sigma$-finite Borel measure $\lambda$ supported
on a compact subset of $\RR^{d+1}$, its potential $U_s ^\lambda\;(0 \le s <d)$ is an
$\alpha$-superharmonic function. Furthermore, Theorem 1.29 in \cite{landkof},
aside from some changes in notation,  states the
following result:

\begin{theorem} \label{dominat}
Suppose $\lambda$ is a
$\sigma$-finite positive Borel measure supported
on a compact subset of $\RR^{d+1}$ whose potential $U_s^\lambda$ is finite $\lambda$-almost
everywhere, and that $f(x)$ is an $\alpha$-superharmonic function. If the inequality
$$U_s^\lambda(x) \le f(x)$$
holds $\lambda$-almost everywhere, then it holds everywhere in $\RR^{d+1}$.
\end{theorem}

Theorem~\ref{dominat} is often called the ``principal of domination" for
$\alpha$-superharmonic functions.

\begin{lemma} \label{exfield}
Assume $d-1 \le s<d$. Then there exists a constant $C_{d,s}$, independent of $N$, such that
for all $x \in \RR^{d+1}$ with $|x| = 1 + N^{-1/d}$,
\beqn
U_s^{v_N}(x) \ge \gamma_{d,s}-C_{d,s}N^{-1+s/d}. \label{keyestimate}
\eeqn
\end{lemma}
\proof By Lemma~\ref{kuijlaars}, we have
$$U_s^{v_N}(x) \ge \gamma_{d,s}-C_{d,s}N^{-1+s/d}, \quad |x|=1.$$
Note that $U_s^\mu(x)=\gamma_{d,s}, \; x \in S^d$, so we can rewrite
the above inequality as
$$U_s^{v_N}(x) \ge U_s^\mu(x)(1-C'_{d,s}N^{-1+s/d}), \quad |x|=1.$$
Both measures
$\mu$ and $v_N$ are supported on $S^d$. Since $U_s^{v_N}$
is an $\alpha$-superharmonic function, we can therefore use Theorem~\ref{dominat} to
obtain for $N$ sufficiently large,
$$U_s^{v_N}(x) \ge U_s^\mu(x)(1-C'_{d,s}N^{-1+s/d}), \quad
 x \in \RR^{d+1}.$$
We then use  Lemma~\ref{le3} to get for $ |x|=1+N^{-1/d}$,
\beqns U_s^{v_N}(x)
&\ge& (\gamma_{d,s}-C''_{d,s}N^{-1+s/d})(1-C'_{d,s}N^{-1+s/d}),\eeqns
which yields the desired estimate. \endproof
\begin{lemma} \label{upperbound}
For every fixed $i, (1 \le i \le N)$, we have, for $0<s<d$,
$$N^{-1}\sum_{j:j \ne i}|x_i-x_j|^{-s}\le \gamma_{d,s}.$$
\end{lemma}
\proof For every fixed $i, \; (1 \le i \le N)$, since the function
$x \mapsto \sum_{j: j  \ne  i}|x-x_j|^{-s}$ reaches its minimum at $x_i$ on $S^d$, we have
$$N^{-1}\sum_{j: j  \ne  i}|x_i-x_j|^{-s} \le N^{-1}\sum_{j: j  \ne  i}|x-x_j|^{-s}, \quad x \in S^d.$$
Integrating both sides of the above inequality on $S^d$ against the probability
measure $\mu(x)$  yields:
$$N^{-1}\sum_{j: j  \ne  i}|x_i-x_j|^{-s} \le N^{-1}\int_{S^d}
\sum_{j: j  \ne  i}|x-x_j|^{-s}d\mu(x)= {N-1 \over N}\gamma_{d,s},$$
and therefore the desired inequality follows. \endproof

\begin{theorem} \label{domination}
Assume $d-1 \le s <d$. The minimal $s$-energy points are well-separated, i.e.,
there exists a constant
$A_{d,s}>0$, independent of $N$, such that
$$ \min_{i \ne j}|x_i-x_j| \ge A_{d,s}N^{-1/d}.$$
\end{theorem}

\proof Let $\omega_N$ be a minimal $s$-energy configuration, and let
$x_{i_0}, x_{j_0}$ be two points in $\omega_N$ such that
$$|x_{i_0}- x_{j_0}|=\min_{i \ne j}|x_i-x_j|.$$ Using Lemma~\ref{upperbound}
we have
\begin{eqnarray} \nonumber
      \gamma_{d,s} - N^{-1}|x_{i_0}- x_{j_0}|^{-s}
       & \ge &N^{-1}\sum_{j:j \ne i_0} |x_{i_0}- x_{j}|^{-s} - N^{-1}|x_{i_0}- x_{j_0}|^{-s} \\
       \label{estimate}
       & =  &N^{-1}\sum_{j:j \ne i_0,j_0} |x_{i_0}- x_{j}|^{-s}.
\end{eqnarray}
Take $x := (1+N^{-1/d})  x_{i_0}$.
Then $|x_{i_0}-x_j| < |x - x_j|$ for every $j$, and so by
(\ref{estimate}) and Lemma \ref{exfield} we have
\begin{eqnarray*}
      \gamma_{d,s} - N^{-1}|x_{i_0}- x_{j_0}|^{-s}
       & \ge & N^{-1}\sum_{j:j \ne i_0,j_0} |x- x_{j}|^{-s}  \\
       & =   & U_s^{v_N}(x) - N^{-1}|x- x_{i_0}|^{-s} - N^{-1}|x- x_{j_0}|^{-s}\\
      &  \ge &  \gamma_{d,s} - C_{d,s} N^{-1+s/d}- N^{-1}|x- x_{i_0}|^{-s}
              - N^{-1}|x- x_{j_0}|^{-s}.
              \end{eqnarray*}
Since $|x-x_{i_0}| = N^{-1/d}$ and $|x-x_{j_0}| > N^{-1/d}$, it follows that
\beqns \gamma_{d,s} - N^{-1}|x_{i_0}- x_{j_0}|^{-s}
        & \ge &  \gamma_{d,s} - (C_{d,s}+2) N^{-1+s/d}, \eeqns
         which implies that for some constant $A_{d,s}$,
         $$|x_{i_0}-x_{j_0}| \ge A_{d,s}N^{-1/d}. \mendproof$$
If one follows the trail of the constants throughout the proofs, then one is able to quantitatively
estimate the value of the constant $A_{d,s}$ in Theorem \ref{domination}. However, such a process
leads to the piling-up of many Gamma function values, among others. There seems to be no obvious way
to simplify them. Just for curiosity, we numerically estimated the constant
$A_{d,s}$ in Theorem \ref{domination} for the case $d=2$, and $s=1$. Our numerical result yields
$A_{2,1} \ge 0.8709$, putting our estimate of the minimum separation of the corresponding minimal energy points
on $S^2$ at $0.8709/\sqrt{N}$. The result of Habicht and Van der Waerden \cite{hw} for best packing
asserts that the maximum diameter of $N$ non-overlapping congruent circles on $S^2$ is asymptotically
$$ \left({8 \pi \over \sqrt{3}}\right)^{1/2} {1 \over \sqrt{N}}\approx 3.809{1 \over \sqrt{N}}.$$
For more information on best packing on $S^2$, we also refer readers to \cite{cs}, \cite{ks2},
and the references therein.
\medskip

\centerline{\bf Acknowledgment}
We thank the referees for pointing out some oversights and
inaccuracies in the initial version of the paper.

\bigskip
\no Department of Mathematics\hfil\break
\no Katholieke Universiteit Leuven\hfil\break
\no Celestijnenlaan 200 B\hfil\break
\no B-3001 Leuven (Heverlee)\hfil\break
\no BELGIUM\hfil\break
\no E-mail: {\tt arno@wis.kuleuven.ac.be}

\bigskip
\no Center for Constructive Approximation\hfil\break
\no Department of Mathematics \hfil\break
\no Vanderbilt University \hfil\break
\no Nashville, TN 37240, USA\hfil\break
\no E-mail: {\tt esaff@math.vanderbilt.edu}

\bigskip
\no Department of Mathematics \hfil\break \no Missouri
State University\hfil\break \no Springfield, MO 65804, USA
\hfil\break \no E-mail: {\tt xis280f@smsu.edu}

\end{document}